\documentclass[aps,pra,twocolumn,amsmath,amssymb,superscriptaddress,longbibliography]{revtex4-1}

\usepackage{mathrsfs}
\usepackage{graphicx}
\usepackage{amsmath}
\usepackage{amssymb}
\usepackage{amsfonts}
\usepackage{color}
\usepackage{CJK}
\usepackage{appendix}

\usepackage{multirow}
\usepackage{booktabs}
\usepackage{bigstrut}
\usepackage{lineno}

\usepackage{lipsum}

\usepackage[unicode=true,colorlinks=true,citecolor=blue,urlcolor=blue]{hyperref}

\def\hatd#1{\hat{#1}^\dagger}

\def\braket#1#2{\left\langle{{#1}}\mathrel{\left|{\vphantom{{#1}{#2}}}\right.\kern-\nulldelimiterspace}{{#2}}\right\rangle}

\begin{document}

\title{Zero-energy Quantum Many-Body Scar under Emergent Chiral Symmetry and Pseudo Hilbert Space Fragmentation}

\author{Li Zhang}
\affiliation{Institute of Quantum Precision Measurement, State Key Laboratory of Radio Frequency Heterogeneous Integration, Shenzhen University, Shenzhen 518060, China}
\affiliation{College of Physics and Optoelectronic Engineering, Shenzhen University, Shenzhen 518060, China}

\author{Yongguan Ke}
\affiliation{Institute of Quantum Precision Measurement, State Key Laboratory of Radio Frequency Heterogeneous Integration, Shenzhen University, Shenzhen 518060, China}
\affiliation{College of Physics and Optoelectronic Engineering, Shenzhen University, Shenzhen 518060, China}

\author{Chaohong Lee}
\altaffiliation{Email: chleecn@szu.edu.cn}
\affiliation{Institute of Quantum Precision Measurement, State Key Laboratory of Radio Frequency Heterogeneous Integration, Shenzhen University, Shenzhen 518060, China}
\affiliation{College of Physics and Optoelectronic Engineering, Shenzhen University, Shenzhen 518060, China}
\affiliation{Quantum Science Center of Guangdong-Hongkong-Macao Greater Bay Area (Guangdong), Shenzhen 518045, China}

\begin{abstract}
Hilbert space fragmentation (HSF) is a mechanism for generating quantum many-body scar (QMBS), which provides a route to weakly break ergodicity.
Many scarred systems possess an exponentially large number of zero-energy states due to the chiral symmetry induced bipartition of the Hilbert space.
In this work, we study the QMBS phenomenology under the interplay between the chiral symmetry and pseudo HSF, where the Hilbert space is approximately fragmented into different blocks.
We consider a model of tilted chain of interacting spinless fermions with periodically varying tunneling strength.
At small tunneling strength, we analytically derive the resonance conditions under which the system is described by an effective model with chiral symmetry and pseudo HSF.
We find that the interplay between the two gives rise to a highly localized zero-energy QMBS when the particle number is even.
This zero-energy QMBS induces an unusual scarred dynamical phenomenon.
Specifically, the fidelity from a simple initial state oscillates around a finite fixed value without decaying, instead of showing the typical decaying collapse and revival observed when the particle number is odd and in common scarred systems.
We show that the signature of the unusual scarred dynamical behaviour can also be detected in the original driven system by measuring local observables.
Our findings enrich the scar phenomenon and deepen the understanding of the relation between Hilbert space structure and QMBS. \\

\end{abstract}


\maketitle

\section{Introduction}

Generic isolated quantum many-body systems are expected to thermalize from arbitrary nonequilibrium initial states via eigenstates thermalization hypothesis (ETH), which states that every bulk eigenstate is indistinguishable from the microscopic ensemble with the corresponding energy as far as local observables are concerned~\cite{Deutsch1991,Srednicki1994,Srednicki1999,Rigol2008,Deutsch2018}.
While the ETH is confirmed in various non-integrable systems, there are systems in which it is violated.
Well-known examples are quantum integrable systems and many-body localized systems, which break the ETH strongly due to an extensive set of conserved quantities~\cite{Rigol2007,Steinigeweg2013,Essler2016,Calabrese2016,Vidmar2016,Gornyi2005,Basko2006,Nandkishore2015,Abanin2019}.

Recently, quantum many-body scars (QMBSs), the nonthermal eigenstates embedded in the thermal spectrum, have attracted much interest as a weak violation of the ETH~\cite{Bernien2017,Turner2018,Turner2018_2,Ho2019,Michailidis2020,Turner2021,Serbyn2021,Chandran2023}.
Although the QMBSs only constitute a vanishing fraction of the eigenspectrum, they can leave an imprint on the dynamics.
While most initial states thermalize quickly after quench, the initial states having high overlap with the QMBSs show slow thermalization and persistent collapse and revivals~\cite{Bernien2017,Turner2018,Turner2018_2}.
One of the mechanisms for generating QMBSs is the (pseudo) Hilbert space fragmentation (HSF), where the Hilbert space is (approximately) decoupled to small subspaces and large thermal subspaces~\cite{Sala2020,Zhao2020,Hudomal2020,Desaules2021,ZhangP2023,Guo2023,Serbyn2021}.
When evolving from an extremal state in a small subspace, the (weak connection) disconnection between the subspace and the rest part of the Hilbert space will protect the wave function from leaking out of that subspace and slows down thermalization.
The QMBSs have been explored widely across various settings, including both static and Floquet systems~\cite{Turner2018,Turner2018_2,Moudgalya2018,Moudgalya2018_2,Schecter2019,Shibata2020,Lin2020,Surace2021,Karle2021,Guo2023,Udupa2023,Wang2024,Bull2019,Hudomal2020,Desaules2021,Banerjee2021,Biswas2022,Sau2024,Zhao2020,Mukherjee2020,Mukherjee2020_2,Mizuta2020,Iadecola2020,Sugiura2021,Rozon2022,Park2023}.
Important experimental progress has also been made in platforms involving Rydberg atoms, ultracold atoms in optical lattices, and superconducting circuits~\cite{Bernien2017,Bluvstein2021,Jepsen2022,Su2023,ZhangP2023}.

%
Interestingly, a large fraction of the scarred models exhibit exponentially many (in size) zero-energy eigenstates in the middle of the spectrum~\cite{Turner2018,Turner2018_2,Lin2020,Surace2021,Buijsman2022,Hudomal2020,Banerjee2021,Biswas2022,Sau2024,Karle2021,Udupa2023,Brighi2023,Brighi2024}, most of which are protected by the intertwining of the chiral and spatial inversion symmetries~\cite{Schecter2018}.
Some works also show that the chiral symmetry along can lead to a large number of zero-energy eigenstates~\cite{Hudomal2020,Brighi2023,Brighi2024}.
While the zero-energy subspace is expected to satisfy the ETH~\cite{Schecter2018}, recent works demonstrate that low-entangled zero-energy eigenstates as QMBSs can exist through appropriate linear combinations, both analytically~\cite{Lin2019,Lin2020,Surace2021,Karle2021,Biswas2022,Udupa2023,Sau2024} and numerically~\cite{Banerjee2021,Karle2021,Biswas2022,Udupa2023,Sau2024}.
The presence of the zero-energy states originates from that the Hilbert space can be divided into bipartite subspaces which are induced by the chiral symmetry, and the number of zero-energy states are bounded from below by the difference of the dimension of the two subspaces~\cite{Inui1994}.
An interesting question then naturally arises that how the interplay between the bipartite structure and the (pseudo) fragmentation of the Hilbert space influences the scar phenomenology.

In this article, we propose a model with chiral symmetry and pseudo HSF, and explore the scar phenomenology under their interplay.
We consider a tilted chain of interacting spinless fermions with periodically varying tunneling strength at half-filling.
In the limit of small tunneling strength, the system is described by an effective Floquet Hamiltonian with on-site potential and density-dependent tunnelings.
We analytically derive the resonance conditions, under which the on-site potential vanishes and the tunneling rate for one tunneling process is always relatively smaller than that of the others.
These lead to an emergence of the chiral symmetry of the effective Hamiltonian and a pseudo fragmentation of the Hilbert space.
According to the chiral symmetry, we analytically derive the lower bound of the number of the zero-energy states and find a remarkable even-odd effect.
Specifically, when the particle number $N$ is even, the lower bound increases exponentially with the system size; when $N$ is odd, it equals 0.
Numerically, we find that the lower bound is always satisfied.
Furthermore, we find that when $N$ is even, the zero-energy subspace supports a localized scar state under the interplay between the chiral symmetry and the pseudo HSF.
This localized zero-energy scar leads to an unusual scarred dynamic phenomenon, where the fidelity shows oscillation around a fixed value without decay, in contrast to the usual decaying collapse and revival in other common scarred systems.
We show that this unusual scarred dynamics can be captured by the original driven model in the finite but not limiting parameter regime.

\section{Model and effective Hamiltonian}\label{Sec2}
We consider an ensemble of interacting spinless fermions in a one-dimensional (1D) tilted lattice under periodic driving, which is described by the Hamiltonian
\begin{eqnarray}\label{Eq.Ham}
\hat H(t)&=&\hat H_{\mathrm{on}}+\left[1+u(t)\right]\hat H_J,\nonumber\\
\hat H_{\mathrm{on}}&=&U\sum_{j=1}^{L-1}\hat n_j\hat n_{j+1}-g\sum_{j=1}^{L}j\hat n_j, \nonumber\\
\hat H_J&=&J\sum_{j=1}^{L-1}(\hatd c_j\hat c_{j+1}+\hatd c_{j+1}\hat c_j).
\end{eqnarray}
Here, $\hatd c_j (\hat c_j)$ creates (annihilates) a fermion at site $j$, and $\hat n_j=\hatd c_j\hat c_j$ is the particle number operator.
The parameters $J$, $U$ and $g$ are the nearest-neighbor tunneling strength, the nearest-neighbor interaction strength and the tilting field strength, respectively.
$u(t)=-u$ and $+u$ in the first and second half of each period $T$, respectively. $\omega=2\pi/T$ is the driving frequency.
$L$ is the total number of lattice sites.
We note that Hamiltonian~(\ref{Eq.Ham}) does not possess the chiral symmetry.
In the following study, we consider open boundary condition and focus on the half-filling sector with particle number $N=L/2$.
We consider $g,U>0$, set $\hbar=1$ and set the energy unit as $J=1$.

The dynamics of the system at stroboscopic times is governed by the Floquet Hamiltonian
\begin{equation}
\hat H_F=\frac{i}{T}\ln\hat F,
\end{equation}
where the Floquet operator $\hat F=\mathcal Te^{\int_0^T\hat H(t)dt}$ is the unitary evolution operator over one period.
Extracting an explicit exact expression for $\hat H_F$ is difficult, because $\hat H_{\mathrm{on}}$ and $\hat H_J$ do not commute with each other.
When $J, uJ\ll g,U,\omega/2\pi$, and $g$ and $U$ are comparable, we can treat $[1+u(t)]\hat H_J$ as a perturbation to $\hat H_{\mathrm{on}}$ and derive an effective Floquet Hamiltonian up to first order (see Appendix~\ref{AppA}) through the time-dependent perturbation theory~\cite{Soori2010,Sen2021,Zhang2024}.
The effective Hamiltonian can be understood as a result of driving-assisted tunnelings and is not chiral symmetric generally.
However, if all the tunneling processes are resonant, the effective Hamiltonian possesses the chiral symmetry.

A tunneling process is resonant if the frequency provided by the driving can compensate the corresponding energy barrier.
We can derive the full resonance condition under which all the tunneling processes are resonant (see Appendix~\ref{AppB}),
\begin{eqnarray}\label{Eq.frescondition}
\begin{cases}
\frac{U}{g}=1+\frac{2k_1+1}{2k_2+1} \\
g=(2k_2+1)\omega
\end{cases}
\mathrm{or} \
\begin{cases}
\frac{U}{g}=1-\frac{2k_1+1}{2k_2+1} \ \mathrm{with} \ k_2>k_1 \\
g=(2k_2+1)\omega,
\end{cases}
\end{eqnarray}
where $k_1$ and $k_2$ are non-negative integers.
In this condition, the first-order effective Floquet Hamiltonian reads
\begin{eqnarray}\label{Eq.Hameff_fres}
\hat H^{(1)}_{F,\mathrm{r}}&=&\frac{2iuJ}{(1-U/g)(2k_2+1)\pi}\sum_j\hat{\mathcal P}^{(|g-U|)}_{j-1,j+2}(\hatd c_{j+1}\hat c_j-h.c.) \nonumber\\
&+&\frac{2iuJ}{(2k_2+1)\pi}\sum_j\hat{\mathcal P}^{(g)}_{j-1,j+2}(\hatd c_{j+1}\hat c_j-h.c.) \nonumber\\
&+&\frac{2iuJ}{(1+U/g)(2k_2+1)\pi}\sum_j\hat{\mathcal P}^{(g+U)}_{j-1,j+2}(\hatd c_{j+1}\hat c_j-h.c.),\nonumber\\
&&
\end{eqnarray}
where the projectors
\begin{eqnarray}\label{Eq.Projctors}
\hat{\mathcal P}_{j-1,j+2}^{(|g-U|)}&=&\hat n_{j+2}(1-\hat n_{j-1}),\nonumber\\
\hat{\mathcal P}_{j-1,j+2}^{(g)}&=&1-(\hat n_{j-1}-\hat n_{j+2})^2, \nonumber\\
\hat{\mathcal P}_{j-1,j+2}^{(g+U)}&=&\hat n_{j-1}\left(1-\hat n_{j+2}\right).
\end{eqnarray}
The three terms in Hamiltonian~(\ref{Eq.Hameff_fres}) represent the tunneling processes characterized by energy barrier $|\Delta|=|g-U|$, $g$ and $g+U$, respectively.
It is obvious that $\hat{\mathcal P}_{j-1,j+2}^{(|g-U|)}+\hat{\mathcal P}_{j-1,j+2}^{(g)}+\hat{\mathcal P}_{j-1,j+2}^{(g+U)}=\hat I$, which means that the Hamiltonian~(\ref{Eq.Hameff_fres}) contains all the nearest-neighboring tunneling processes and is free from dynamical constraints.

\section{Emergent chiral symmetry and pseudo Hilbert space fragmentation}\label{Sec3}
\begin{figure}[!b]
\includegraphics[width=1\columnwidth]{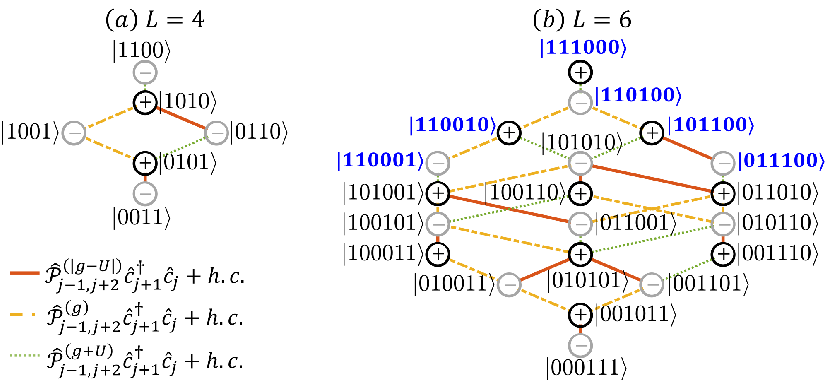}
  \caption{\label{Fig_graph}
  The Hilbert space graph of Hamiltonian~(\ref{Eq.Hameff_fres}) at (a) $L=4$ and (b) $L=6$.
  The chiral symmetric ($+$) and anti-symmetric ($-$) Fock states $|\mathbf n\rangle=|n_1n_2\cdots n_L\rangle$ ($n_j=0$, 1 is the number of particles at site $j$) are denoted by black and grey circles, respectively.
  The tunneling processes with $|\Delta|=|g-U|$, $g$ and $g+U$ are denoted by solid red, dash-dotted yellow, and green dotted lines, respectively.
  The Fock states in bold blue label the states in the tower.
  }
\end{figure}

There are two key properties of Hamiltonian~(\ref{Eq.Hameff_fres}) that will influence the structure of the Hilbert space and potentially the dynamical behaviour.
First, there emerges a chiral symmetry of Hamiltonian~(\ref{Eq.Hameff_fres}) with a chiral operator $\hat{\mathcal C}=(-1)^{\sum_jj\hat n_j}$ anticommuting with $\hat H^{(1)}_{F,\mathrm{r}}$, which will cause the Hilbert space to be bipartite.
Second, the tunneling strength for the process with $|\Delta|=g+U$ is always smaller than that of the other two, which will makes the Hilbert space approximately fragmented.
In the following, we elucidate the two points by considering the Hilbert space graph of Hamiltonian~(\ref{Eq.Hameff_fres}), where each vertex corresponds to a Fock state and two vertices are connected by an edge if the corresponding Hamiltonian matrix element is nonzero.

The chiral symmetry implies that the Hamiltonian matrix elements $\langle \mathbf n'|\hat H_{F,\mathrm r}^{(1)}|\mathbf n\rangle$ between the Fock states are nonzero only when the Fock states $|\mathbf n'\rangle$ and $|\mathbf n\rangle$ belong to different chiral subspaces.
That is, the graph has a bipartite structure with the even/odd sublattices corresponding to the chiral symmetric/antisymmetric subspaces, see Fig.~\ref{Fig_graph}, where the circled $+/-$ denote the chiral symmetric/antisymmetric Fock states.
According to Ref.\cite{Inui1994}, such a bipartite Hilbert space structure will result in at least $|N_+-N_-|$ zero-energy eigenstates distributing only on the larger subspace, where $N_+$ and $N_-$ are the dimensions of the chiral symmetric and anti-symmetric subspaces, respectively.
For our system, we can prove that (see Appendix~\ref{AppC})
\begin{equation}\label{Eq.SubSpaceDimDiff}
|N_+-N_-|= \begin{cases}
0, & \text{if } N \text{ is odd}, \\
\frac{N!}{(N/2)!(N/2)!}, & \text{if } N \text{ is even},
\end{cases}
\end{equation}
which means that there will be at least exponentially many zero-energy eigenstates when $N$ is even.


If we ignore the tunneling process with $|\Delta|=g+U$, the Hilbert space will be fragmented into different blocks.
The relatively smaller tunneling strength leads to a pseudo HSF, where the blocks are connected weakly through this tunneling process.
To show it explicitly, in Fig.~\ref{Fig_graph}, we label the tunneling processes by different lines with the thinnest dotted lines depicting the one with $|\Delta|=g+U$.
When $N>2$, due to the pseudo fragmentation, there exists a tower-like structure which is weakly connected to the rest part, see the Fock states in bold blue in Fig.~\ref{Fig_graph}(b).
The tower contains a pinnacle state $|\mathrm{tp}\rangle=|(1\cdots1)^N(0\cdots0)^N\rangle$ and $L-1$ eave states $|\mathrm{te}_{\mathrm p,q}\rangle=|(1\cdots1)^{N-1}(0\cdots0)^q1(0\cdots0)^{N-q}\rangle$ and $|\mathrm{te}_{\mathrm h,q}\rangle=|(1\cdots1)^{N-q}0(1\cdots1)^q(0\cdots0)^{N-1}\rangle$, where $q=1,2,\cdots, N$ and $(1\cdots1)^k$ and $(0\cdots0)^k$ denote $k$ contiguous occupied and empty sites, respectively.
$|\mathrm{tp}\rangle$ connects to $|\mathrm{te}_{\mathrm{p/h},1}\rangle$ weakly by breaking the domain at the boundary of the two halfs of the chain, and a pair of particle-hold is created at the boundary.
$|\mathrm{te}_{\mathrm{p/h},q}\rangle$ with $q>1$ is a result of the particle/hole moving through the right/left $N-1$ empty/occupied sites until reaching the boundary.
The eave states connect to the states outside the tower weakly by breaking a domain $\cdots 1100\cdots$ while $|\mathrm{tp}\rangle$ does not directly connect to the rest of the Hilbert space, which will suppress the information leakage outside the tower from $|\mathrm{tp}\rangle$.
Moreover, when $N$ is even, $|\mathrm{tp}\rangle$ is always in the larger chiral subspace (see Appendix~\ref{AppC}), which means that $|\mathrm{tp}\rangle$ may contribute to the zero-energy states.
We note that, due to different tunneling form, the tower structure is different from the hypercube geometry studied in Refs.~\cite{Desaules2021} and~\cite{Guo2023}, which represents spin paramagnets by viewing two lattice sites as cells.
In the following, we show that when $N$ is even, the interplay between the bipartite structure and the pseudo fragmentation of the Hilbert space leads to unusual scarred dynamics which are related to a zero-energy QBMS localized in the Hilbert space.

%
%
%
%
\section{The phenomenology of the zero-energy QMBS}\label{Sec4}
\subsection{Unusual scarred dynamics}

\begin{figure}[!b]
\includegraphics[width=1\columnwidth]{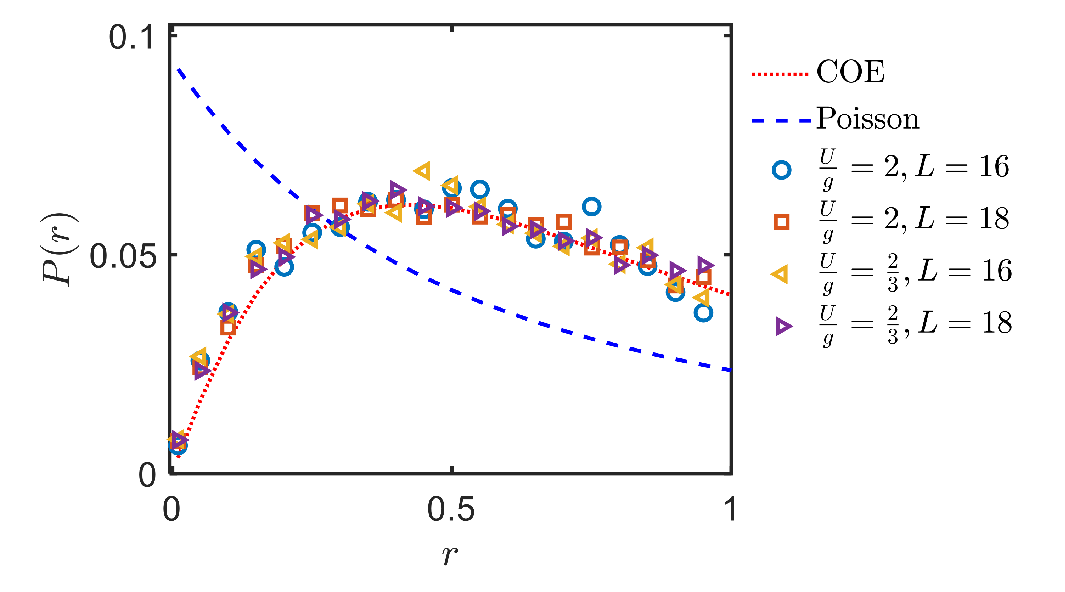}
  \caption{\label{Fig_r}
  The normalized distribution $P(r)$ of the quasienergy gap ratio at $(U/g,L)=(2,16)$ (blue circles), (2,18) (red squares), $(2/3,16)$ (yellow left-pointing triangles) and $(2/3,18)$ (purple right-pointing triangles).
  The red dotted line and the blue dashed line denote the COE distribution and the Poisson distribution, respectively.
  The other parameters are $g=50$, $\omega=g$ and $u=0.5$.
  }
\end{figure}

Before studying the dynamics, we show that the Hamiltonian~\eqref{Eq.Hameff_fres} is nonintegrable, from which one expects the dynamics to be thermal.
We analyze the statistic of the Floquet quasienergies $\{\epsilon_\alpha\}$, which are obtained by folding the eigenenergies of Hamiltonian~\eqref{Eq.Hameff_fres} into one Floquet-Brillouin zone $[-\omega/2,\omega/2)$.
We calculate the distribution $P(r)$ of the consecutive quasienergy gap ratio $r_\alpha=\min{\left(\delta_{\alpha+1}/\delta_{\alpha},\delta_{\alpha}/\delta_{\alpha+1}\right)}$ with $\delta_{\alpha}=\epsilon_{\alpha}-\epsilon_{\alpha-1}$.
Since Hamiltonian~\eqref{Eq.Hameff_fres} is chiral symmetric and thus its spectrum is symmetric about 0, we only consider the positive part of the spectrum.
We find that $P(r)$ satisfies the Wigner-Dyson distribution of circular orthogonal ensemble (COE) $P_{\mathrm{COE}}(r)$~\cite{D'Alessio2014,Ponte2015_2}, see Fig.~\ref{Fig_r} for $(U/g,g/\omega)=(2,1)$ and $(2/3,1)$ at $L=16$ and 18.
This means that Hamiltonian~\eqref{Eq.Hameff_fres} is nonintegrable and one usually expects thermal dynamics.

\begin{figure}[!b]
\includegraphics[width=1\columnwidth]{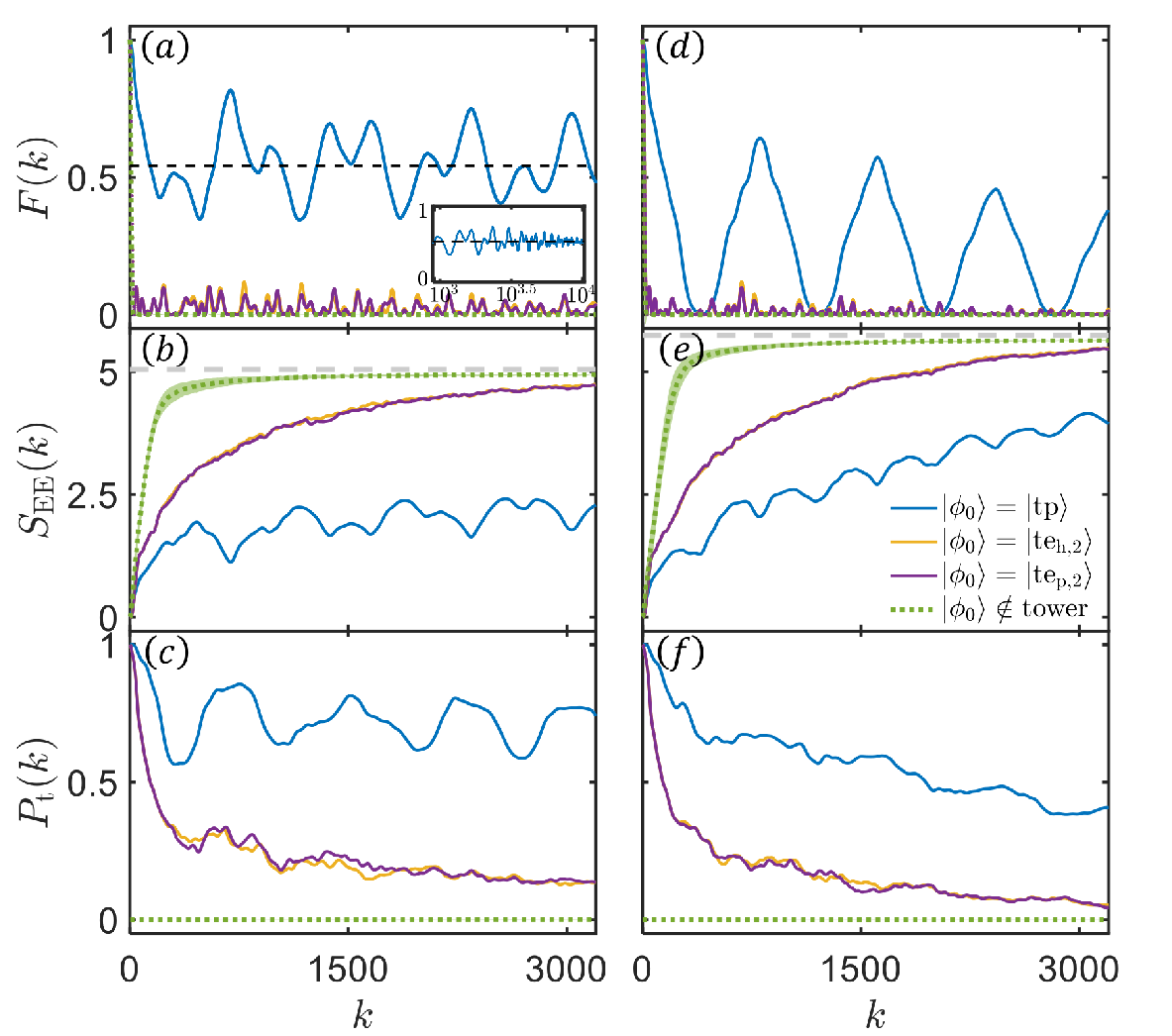}
  \caption{\label{Fig_dynamics}
  Dynamics after quenching from $|\mathrm{tp}\rangle$, $|\mathrm{te}_{\mathrm{p/h},2}\rangle$ and random Fock states outside the tower at system sizes $L=16$ (left column) and $L=18$ (right column).
  $|\phi_0\rangle\notin \mathrm{tower}$ denotes the average over 10 initial random Fock states outside the tower, and the shading denotes the standard deviation.
  Top row: $F(k)$ versus $k$.
  The black dashed line in (a) denotes $P_{0,\mathrm{tp}}^2$.
  Inset in (a): $F(k)$ for longer time when quenching from $|\mathrm{tp}\rangle$.
  Middle row: $S_{\mathrm{EE}}(k)$ versus $k$.
  The gray dashed lines denote the Page value $S_p$ of the systems.
  Bottom row: $P_{\mathrm t}(k)$ versus $k$.
  For $F(k)$ and $P_{\mathrm t}(k)$, the standard deviations are too small to be visible.
  The other parameters are $U/g=2$, $g/\omega=1$, $g=50$ and $u=0.5$.
  }
\end{figure}

We now study the dynamical evolution from different initial states $|\phi_0\rangle$, including $|\mathrm{tp}\rangle$, two eave states $|\mathrm{te}_{\mathrm{p/h},2}\rangle$ and ten randomly chosen Fock states outside the tower.
We simulate the dynamical evolution at $U/g=2, g/\omega=1, g=50$, $u=0.5$ and $L=16$ and 18 through the exact diagnalization (ED) method.
In Fig.~\ref{Fig_dynamics}(a) and (d), we plot the evolution of the fidelity $F(k)=|\langle \phi_0|\exp[-i\hat H^{(1)}_{F,\mathrm{r}}kT]|\phi_0 \rangle|^2$.
For $|\mathrm{te}_{\mathrm{p/h},2}\rangle$ and the initial states outside the tower, the fidelity shows the expected thermal dynamics with $F(k)$ decaying to 0 rapidly for both system sizes.
For $|\mathrm{tp}\rangle$, the fidelity $F(k)$ shows obvious non-thermal dynamics which is different for even and odd $N$.
While $F(k)$ displays the typical scar signature of decaying coherent revivals~\cite{Turner2018_2,Hudomal2020,Desaules2021,ZhangP2023,Guo2023} when $L=18$, it shows unusual dynamics of oscillating around a fixed value without decaying when $L=16$.
The dependence of the scarred dynamics on the odevity of the particle number is also observed for other system sizes, see Appendix~\ref{AppD}.
Fig.~\ref{Fig_dynamics}(b) and (e) show the evolution of the half-chain von Neumann entanglement entropy (EE) $S_{\mathrm{EE}}(k)$, which is defined as $S_{\mathrm{EE}}(k)=-\mathrm{Tr}[\hat \rho_{\mathrm R}(k)\ln \hat \rho_{R}(k)]$ with $\hat \rho_{\mathrm{R}}(k)$ being the reduced density matrix of the right half chain at time $kT$.
For $|\mathrm{te}_{\mathrm{p/h},2}\rangle$ and the initial states outside the tower, $S_{\mathrm{EE}}(k)$ rapidly grows and reaches the Page value $S_p\approx L/2\ln2-1/2$~\cite{Page1993}.
For $|\mathrm{tp}\rangle$, the EE shows much slower growth, which further illustrate the scarred dynamics from $|\mathrm{tp}\rangle$.

\begin{figure}[!t]
\includegraphics[width=1\columnwidth]{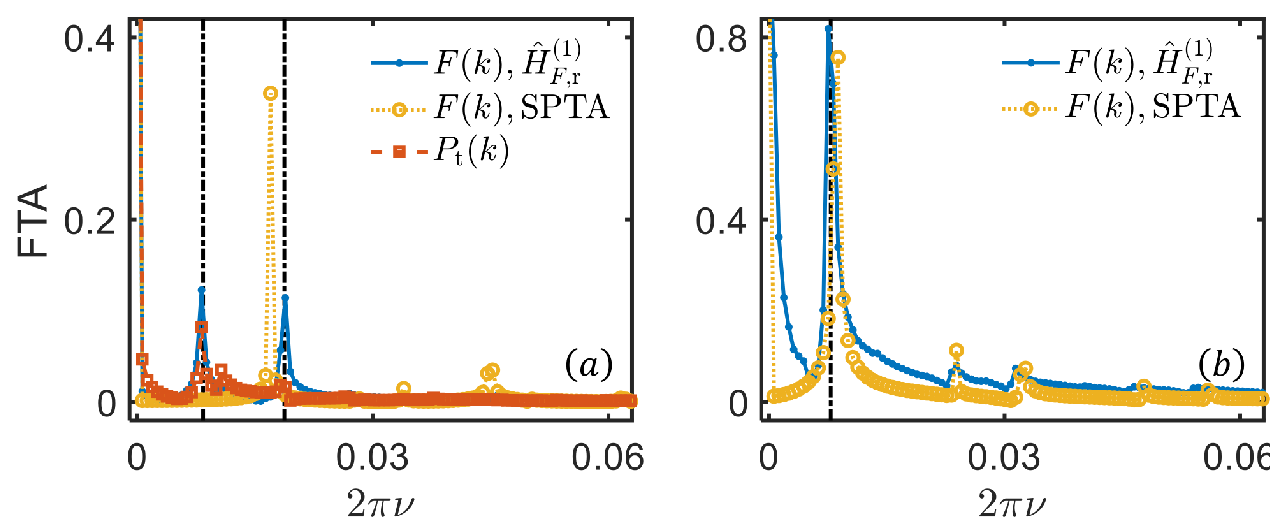}
  \caption{\label{Fig_FFT}
  The FTA (normalized to the maximum value) of $F(k)$ before and after the SPTA when evolving from the initial state $|\mathrm{tp}\rangle$ at (a) $L=16$ and (b) $L=18$.
  In (a), the red dashed line marked by squares denotes the normalized FTA of $P_{\mathrm t}(k)$.
  The vertical black dot-dashed lines denote the frequencies (a) $|\tilde\epsilon_{1(2),L=16}T|$ and (b) $|2\tilde\epsilon_{L=18}T|$.
  }
\end{figure}

The nonthermal dynamics from $|\mathrm{tp}\rangle$ is a result of the suppression of the wave function from leaking out of the tower.
In Fig.~\ref{Fig_dynamics}(c) and (f), we plot the probability of the wave functions to stay in the tower $P_{\mathrm t}(k)=\sum_{\mathbf{n}\in{\mathrm{tower}}}|\langle \mathbf n|\exp[-i\hat H^{(1)}_{F,\mathrm{r}}kT]|\phi_0\rangle|^2$ for $L=16$ and $L=18$, respectively.
When evolving from $|\mathrm{tp}\rangle$, $P_{\mathrm{t}}(k)$ remains high compared to other cases for both system sizes.
Additionally, $P_{\mathrm{t}}(k)$ shows slow decaying for $L=18$ while oscillates around a fixed value for $L=16$, reminiscent of the different scarred dynamics in Fig.~\ref{Fig_dynamics}(a) and (d).
To further illustrate the effect of the tower structure, we make a single particle tower approximation (SPTA), which projects the Hamiltonian~\eqref{Eq.Hameff_fres} into the tower and decouples the tower from the rest of the graph.
Fig.~\ref{Fig_FFT} shows the Fourier transformation amplitude (FTA) of $F(k)$ when evolving from $|\mathrm{tp}\rangle$ before and after the SPTA.
For $L=18$, the peak location of the FTA after the SPTA well matches that evolved under Hamiltonian~\eqref{Eq.Hameff_fres}, which implies that the wave function revival is a result of the interference effect of a single particle walking in the tower of finite length.
For $L=16$, the fidelity evolved under Hamiltonian~\eqref{Eq.Hameff_fres} shows two oscillation frequencies.
The high frequency oscillation is well captured by the SPTA.
To analyze the low frequency oscillation, we make the Fourier transformation of $P_{\mathrm t}(k)$.
We find that the peak location of the FTA of $P_{\mathrm t}(k)$ well matches the low oscillation frequency of $F(k)$, which means that the low frequency oscillation comes from that part of the wave function bounces between the tower and the rest of the graph and deserves further study.
We note that the scarred dynamics might not persist in the thermodynamic limit, possibly for the reason that the size of the tower and the number of the edges connecting the tower and the thermal part of the graph increase with $L$, see Appendix~\ref{AppD}.

\subsection{Zero-energy quantum many-body scar}
We now explain the mechanism for the unusual scarred dynamical phenomenon for even $N$.
For this purpose, we study the properties of the Floquet states $|\psi_\alpha\rangle$, which are the eigenstates of the Hamiltonian~\eqref{Eq.Hameff_fres}.
We first derive the expression for $F(k)$ from any initial Fock state.
Expanding the Floquet states in the Fock basis as $|\psi_\alpha\rangle=\sum_{\mathbf{n}}\psi_{\mathbf{n},\alpha}|\mathbf n\rangle$ and utilizing the chiral symmetry, the fidelity from an initial Fock state $|\mathbf n_0\rangle$ is derived as
\begin{eqnarray}\label{Eq.Fidelity}
F(k)&=&P_{0,\mathbf{n_0}}^2+4P_{0,\mathbf{n_0}}\sum_{\alpha:\epsilon_\alpha>0}|\psi_{\mathbf n_0,\alpha}|^2\cos\left(\epsilon_\alpha kT\right)\nonumber\\
&+&2\sum_{\alpha:\epsilon_{\alpha}>0\atop \beta:\epsilon_{\beta}>0}|\psi_{\mathbf n_0,\alpha}|^2|\psi_{\mathbf n_0,\beta}|^2\cos\left[\big(\epsilon_\alpha+\epsilon_\beta\big)kT\right] \nonumber\\
&+&2\sum_{\alpha:\epsilon_{\alpha}>0\atop \beta:\epsilon_{\beta}>0}|\psi_{\mathbf n_0,\alpha}|^2|\psi_{\mathbf n_0,\beta}|^2\cos\left[\big(\epsilon_\alpha-\epsilon_\beta\big)kT\right],
\end{eqnarray}
where $P_{0,\mathbf{n_0}}=\sum_{\alpha:\epsilon_\alpha=0}|\psi_{\mathbf n_0,\alpha}|^2$ is the projection of $|\mathbf n_0\rangle$ onto the zero-energy subspace.
$F(k)$ will oscillate around $\bar F=P_{0,\mathbf{n_0}}^2+2\sum_{\alpha:\epsilon_\alpha>0}|\psi_{\mathbf n_0,\alpha}|^4$, with frequencies determined by the quasienergies of the Floquet states which have high overlap with $|\mathbf n_0\rangle$.
We numerically diagonalize the Hamiltonian~\eqref{Eq.Hameff_fres}, and find that the number of zero-energy states satisfies the lower bound Eq.~\eqref{Eq.SubSpaceDimDiff}, at least for the system sizes $L\leq 20$ accessible numerically, which means that $P_{0,\mathbf{n_0}}=0$ for odd $N$.

\begin{figure}[!htbp]
\includegraphics[width=1\columnwidth]{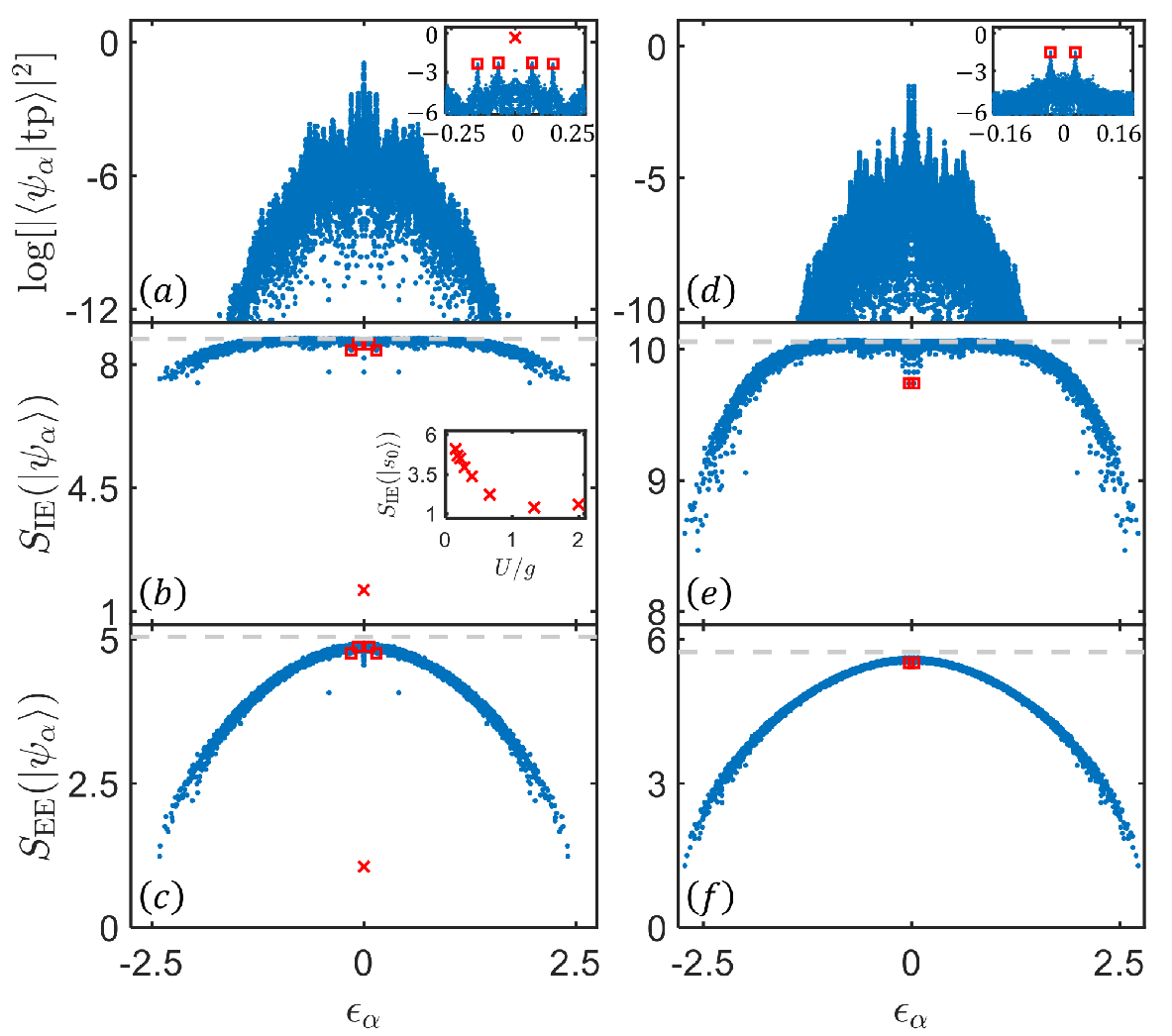}
  \caption{\label{Fig_S_IE}
  The properties of the Floquet states at $U/g=2$, $g/\omega=1$, $g=50$, $u=0.5$ and system sizes $L=16$ (left column) and $L=18$ (right column).
  Top row: the semi-log plot of the projection $|\langle \psi_\alpha|\mathrm{tp}\rangle|^2$ versus $\epsilon_\alpha$.
  Insets: the detail around the vicinity of the peaks.
  In the inset of (a), the projection onto the zero-energy subspace is plotted in the orthogonal basis consisting $|s_0\rangle$.
  Middle row: $S_{\mathrm{IE}}(|\psi_\alpha\rangle)$ versus $\epsilon_\alpha$.
  The grey dashed lines denote the value of IE predicted by the COE statistic.
  Inset in (b): $S_{\mathrm{IE}}(|s_0\rangle)$ at $U/g=2$, $4/3$ and $2/(2k+3)$ with $k=0,1,2,3,4$.
  Bottom row: $S_{\mathrm{EE}}(|\psi_\alpha\rangle)$ versus $\epsilon_\alpha$.
  The grey dashed lines denote the Page value $S_p$ of EE.
  In all plots, the blue dots are the \textcolor{blue}{raw} results from the ED, the red cross stands for the zero-energy QMBS \textcolor{blue}{$|s_0\rangle$} and the red squares denote the non-zero energy Floquet states having relatively high overlap with $|\mathrm{tp}\rangle$.
  }
\end{figure}

In Fig.~\ref{Fig_S_IE}, we show the properties of the Floquet states for $L=16$ and 18 at $U/g=2,g/\omega=1,g=50$ and $u=0.5$.
We calculate the projection of $|\mathrm{tp}\rangle$ onto the Floquet states, the half-chain von Neumann EE $S_{\mathrm{EE}}(|\psi_\alpha\rangle)$ and the Shannon information entropy (IE).
The Shannon IE is defined as $S_{\mathrm{IE}}(|\psi_\alpha\rangle)=-\sum_{\mathbf n}|\langle \mathbf n|\psi_\alpha\rangle|^2\ln|\langle \mathbf n|\psi_\alpha\rangle|^2$, which measures the degree of localization of $|\psi_\alpha\rangle$ in the Fock space~\cite{D'Alessio2014,Santos2010,Santos2010_2}.
If $|\psi_\alpha\rangle$ distributes uniformly among the Fock space, $S_{\mathrm{IE}}(|\psi_\alpha\rangle)$ tends to the COE statistic predicted value $\ln(0.48\mathcal D)$, where $\mathcal D$ is the Hilbert space dimension~\cite{D'Alessio2014,Santos2010,Santos2010_2}.
If $|\psi_\alpha\rangle$ is localized in the Fock space, $S_{\mathrm{IE}}(|\psi_\alpha\rangle)$ has much smaller value.

For $L=18$, $|\mathrm{tp}\rangle$ is relatively high overlapped with the Floquet states around the quasienergies $\tilde \epsilon_{L=18}\approx \pm 0.031$, see the red squares in the inset of Fig.~\ref{Fig_S_IE}(d).
These special states account for the oscillation of the fidelity, see Fig.~\ref{Fig_FFT}(b), where the separation $2|\tilde \epsilon_{L=18}T|$ well matches the oscillation frequency of the fidelity.
The IE has high values closing to the COE predicted value, see Fig.~\ref{Fig_S_IE}(e), which means that the Floquet states are delocalized among the Fock space.
The absence of the zero-energy states and the delocalization of the Floquet states leads to the exponentially small value $\bar F\sim 1/\sqrt{\mathcal D}$.
The EE shows a thermal band and the special states do not deviate from the band, see Fig.~\ref{Fig_S_IE}(f).
These special states are examples of thermal QMBSs, which are thermal but lead to nonthermal dynamics from particular initial states~\cite{Evrard2024}.

For $L=16$, $|\mathrm{tp}\rangle$ is highly weighted on the zero-energy subspace, see the peak at zero energy in Fig.~\ref{Fig_S_IE}(a).
The \textcolor{blue}{raw} zero-energy Floquet states obtained numerically are delocalized among the Fock space and their EE do not deviate from the thermal band of EE, see Fig.~\ref{Fig_S_IE}(b) and (c), which means that they are thermal.
However, motivated by the high overlap between $|\mathrm{tp}\rangle$ and the zero-energy subspace, we can naturally construct a zero-energy state by projecting $|\mathrm{tp}\rangle$ onto the zero-energy sbuspace:
\begin{equation}
|s_0\rangle=\frac{1}{\sqrt{P_{0,\mathrm{tp}}}}\sum_{\alpha:\epsilon_\alpha=0}\langle \psi_\alpha|\mathrm{tp}\rangle|\psi_\alpha\rangle,
\end{equation}
where $\sum_{\alpha:\epsilon_\alpha=0}$ means summation over all the raw zero-energy states and $P_{0,\mathrm{tp}}=\sum_{\alpha:\epsilon_{\alpha}=0}|\langle \psi_\alpha|\mathrm{tp}\rangle|^2$ is the weight of $|\mathrm{tp}\rangle$ on the zero-energy subspace.
$|s_0\rangle$ has much lower value of IE because it is localized around $|\mathrm{tp}\rangle$, see the red cross in Fig.~\ref{Fig_S_IE}(b).
Moreover, $|s_0\rangle$ has much lower EE and obviously separates from the thermal band of EE, see Fig.~\ref{Fig_S_IE}(c).
Therefore, it is a nonthermal zero-energy QMBS, in contrast to the thermal QMBSs referred previously.
In addition to $|s_0\rangle$, the initial state $|\mathrm{tp}\rangle$ has slightly high overlap with the Floquet states around the quasienergies $\tilde \epsilon_{1,L=16}\approx\pm0.064$ and $\tilde \epsilon_{2,L=16}\approx\pm0.15$, see the red squares in the inset of Fig.~\ref{Fig_S_IE}(a).
These special states are thermal QMBSs with the IE and EE not deviating from the respective thermal band, see Fig.~\ref{Fig_S_IE}(b) and (c).
Thus, the large value of $\bar F$ is a result of the existence of the highly localized zero-energy QMBS $|s_0\rangle$, see Fig.~\ref{Fig_dynamics}(a), where the black dashed line denotes $P_{0,\mathrm{tp}}^2$.
The oscillation of $F(k)$ is well captured by $|s_0\rangle$ and the thermal QMBSs, with the two frequencies well predicted by $|\tilde \epsilon_{L=16,1}T|$ and $|\tilde \epsilon_{L=16,2}T|$, see the vertical black dot-dashed lines in Fig.~\ref{Fig_FFT}(a).
We note that the highly localized zero-energy QMBS is a result of the interplay between the chiral symmetry and the pseudo HSF.
When tuning $U/g$ to lower value, the three tunneling strengthes become more uniform and $|s_0\rangle$ becomes less localized, see the inset in Fig.~\ref{Fig_S_IE}(b).

\subsection{Probing the unusual scarred dynamics}
\begin{figure}[!t]
\includegraphics[width=1\columnwidth]{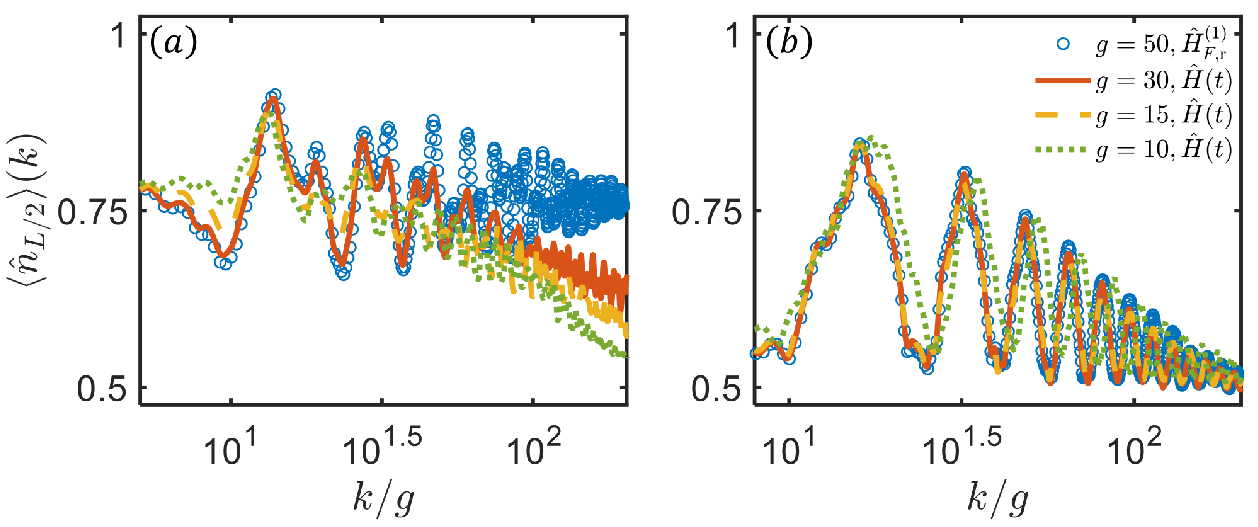}
  \caption{\label{Fig_full_eff}
  The middle site density $\langle \hat n_{L/2}\rangle(k)$ versus $k/g$ under the evolution of $\hat H_{F,\mathrm r}^{(1)}$ (blue circles) and $\hat H(t)$ at different values of $g$ (lines) for (a) $L=16$ and (b) $L=18$.
  The results for $\hat H_{F,\mathrm r}^{(1)}$ are plotted every 20 driving cycles for visibility.
  In (b), the data for $g=15$ and 30 under $\hat H(t)$ almost collapse with each other.
  The other parameters are $U/g=2$, $g/\omega=1$ and $u=0.5$.
  }
\end{figure}

At last, we note that the dependence of the scarred dynamics on the particle odevity can still be probed by the full Hamiltonian~\eqref{Eq.Ham} at large but finite $U$ and $g$ through measruring local observables.
Hamiltonian~\eqref{Eq.Ham} can be simulated by insulating two-component bosonic atoms trapped in a one-dimensional tilted optical lattice~\cite{Liu2020}.
In Fig.~\ref{Fig_full_eff}, we compare the evolution of the density $\langle \hat n_{L/2}\rangle(k)$ governed by the full Hamiltonian~\eqref{Eq.Ham} for $L=16$ and $L=18$, at $U/g=2$, $g/\omega=1$, $u=0.5$ and different values of $g$.
For $L=16$, the density $\langle \hat n_{L/2}\rangle(k)$ oscillates around a fixed value and well mathches that for the the effective model up to driving cycles $k\sim10^{1.5}g$ when $g\gtrsim15$.
When $k\gtrsim 10^{1.5}g$, the density starts to decay to the thermal value 0.5 due to higher order perturbation processes.
For $L=18$, the density shows decaying coherent oscillation and is well consistent with that for the effective model when $g\gtrsim15$.
These mean that the unusual scarred dynamics can be distinguished from the typical scarred dynamics for a range of driving cycles $\sim 10^{1.5}g$ when $g\gtrsim15$.

\section{Summary and discussion}\label{Sec5}
In summary, we have explored the scar phenomenology under the interplay between the chiral symmetry and the pseudo HSF.
We propose an effective model featuring the chiral symmetry and pseudo HSF by leveraging the resonance between a driving field and a titled chain of interacting splinless fermions in the large tilting limit.
We find a remarkable even-odd effect of the particle number on the number of zero-energy states at half-filling.
Specifically, there exist exponentially many zero-energy states when the particle number $N$ is even, whereas no zero-energy states exist when $N$ is odd.
We find that due to the interplay between the chiral symmetry and the pseudo HSF, the zero-energy subspace supports a highly localized zero-energy QMBS.
The localized zero-energy QMBS leads to an unusual scarred dynamic behaviour from a simple initial product state: the fidelity oscillates around a fixed finite value without decaying, in contrast to the typical decaying collapse and revival observed when $N$ is odd and in other scarred models.
The unusual scarred dynamics can also be signaled in the original driven model with large but finite tilting strength by measuring local observables.

The model proposed in our work is unconstrained and the bipartite structure is generic for any 1D lattice of spinless fermions with nearest-neighboring tunneling at half-filling and under open boundary condition.
So, the lower bound of the number of zero-energy states derived in our work is applicable for any such 1D lattice system, regardless of how the tunneling strength depends on the local density or whether the system is a supperlattice.
Therefore, it would be interesting to propose other forms of pseudo HSF and utilize the interplay between it and the zero-energy subspace to protect desired state from thermalizing.
Bipartite structure of Hilbert space graph can also exist in higher dimensional cases. In fact, one can generalize the definition of the chiral operator in the 1D case to higher dimensions straightforwardly. It is easy to see that higher dimensional lattice systems of spinless fermions tunneling between nearest-neighboring sites have chiral symmetry under open boundary condition (for arbitrary number of lattice sites) or periodic boundary condition (for even number of lattice sites along each dimension). In future work, it is interesting to explore the zero-energy subspace and zero-energy QMBSs in such higher-dimensional systems.
Another interesting direction for future work would be to explore the fate of the zero-energy QMBS under chiral symmetry breaking and conserving perturbations, respectively.

\acknowledgments
This study is supported by the National Natural Science Foundation of China (Grants No. 12025509, No. 12305048), Shenzhen Fundamental Research Project (Grants No. JCYJ20230808105009018), the National Key Research and Development Program of China (Grant No. 2022YFA1404104) and Guangdong Provincial Quantum Science Strategic Initiative (GDZX2305006). Y.K. is supported by the National Natural Science Foundation of China (Grant No. 12275365) and the Natural Science Foundation of Guangdong (Grant No. 2023A1515012099).


\appendix


\section{Derivation of the first order effective Floquet Hamiltonian}\label{AppA}
When $J$, $uJ\ll g, U$ and $\omega/2\pi$, we can treat $\hat H_J(t)$ as a perturbation to $\hat H_{\mathrm{on}}$.
By using the Floquet perturbation approach~\cite{Soori2010,Sen2021} and following Ref.~\cite{Zhang2024}, we can derive the effective Floquet Hamiltonian
\begin{equation}
\hat H_F=\frac{i}{T}\ln e^{-i\hat H_{\mathrm{on}}T}+\frac{i}{T}\left[\hat F_1+\hat F_2-\frac{1}{2}\hat F_1^2+\mathcal O(J^3 T^3)\right],
\end{equation}
where the first two orders of Floquet operator
\begin{widetext}
\begin{eqnarray}
\hat F_1&=&-iJ\sum_{\mathbf n,\mathbf l}V_{\mathbf n,\mathbf l}\int_0^Te^{i\Delta_{\mathbf n,\mathbf l}t_1}\left[1+u(t_1)\right]dt_1|\mathbf n\rangle\langle \mathbf l|,\nonumber\\
\hat F_2&=&-J^2\sum_{\mathbf n,\mathbf l}\sum_{\mathbf m_1}V_{\mathbf n,\mathbf m_1}V_{\mathbf m_1,\mathbf l}\int_0^Te^{i\Delta_{\mathbf n,\mathbf m_1}t_1}\left[1+u(t_1)\right]\int_0^{t_1}e^{i\Delta_{\mathbf m_1, \mathbf l}t_2}\left[1+u(t_2)\right]dt_2dt_1|\mathbf n\rangle\langle \mathbf l|.
\end{eqnarray}
\end{widetext}
Here, $|\mathbf n\rangle=|n_0n_1\cdots n_{L-1}\rangle$ with $n_j=0,1$ denote the Fock states, $V_{\mathbf n,\mathbf 1}=\sum_j\langle \mathbf n|\hatd c_j\hat c_{j+1}+\hatd c_{j+1}\hat c_j|\mathbf l\rangle$ is a tunneling matrix element and $\Delta_{\mathbf n,\mathbf l}=E_{\mathbf n}-E_{\mathbf l}$ with $E_{\mathbf n}=\langle \mathbf n|\hat H_{\mathrm{on}}|\mathbf n\rangle$ is the energy barrier for the tunneling process from $|\mathbf l\rangle$ to $|\mathbf n\rangle$.
It is obvious that for a nonzero $V_{\mathbf n,\mathbf l}$, the energy barrier $\Delta_{\mathbf n,\mathbf l}$ can only take one of the six values $\pm|g-U|$, $\pm g$ and $\pm (g+U)$.
Substituting the driving form $u(t)$ to $\hat F_1$, we can obtain the first order effective Floquet Hamiltonian
\begin{widetext}
\begin{eqnarray}\label{Eq.FirstOrderHam}
&&\hat H_F^{(1)}=\frac{i}{T}\ln e^{-i\hat H_{\mathrm{on}}T}+\sum_j\left\{\left[J_1\hat{\mathcal P}_{j-1,j+2}^{(|g-U|)}+J_2\hat{\mathcal P}_{j-1,j+2}^{(g)}+J_3\hat{\mathcal P}_{j-1,j+2}^{(g+U)}\right]\hatd c_j\hat c_{j+1}+h.c.\right\}, \nonumber\\
&&J_1=J\left\{\delta_{g,U}-\frac{i\left(1-\delta_{g,U}\right)}{T(g-U)}\left[e^{i\frac{(g-U)T}{2}}-1\right]\left[u-1+(u+1)e^{i\frac{(g-U)T}{2}}\right]\right\}, \nonumber\\
&&J_2=-\frac{iJ}{Tg}\left(e^{\frac{igT}{2}}-1\right)\left[u-1+\left(u+1\right)e^{\frac{igT}{2}}\right], \nonumber\\
&&J_3=-\frac{iJ}{T(g+U)}\left[e^{\frac{i(g+U)T}{2}}-1\right]\left[u-1+\left(u+1\right)e^{\frac{i(g+U)T}{2}}\right],
\end{eqnarray}
\end{widetext}
where the projectors $\hat{\mathcal P}_{j-1,j+2}^{(|\Delta|)}$ are defined in Eq.~\eqref{Eq.Projctors} with $|\Delta|=|g-U|,g$ and $g+U$.
The first term in Hamiltonian~\eqref{Eq.FirstOrderHam} is an effective on-site potential and the last three projected tunnelings terms represent the driving assisted tunneling processes, which are categorized according to the energy barrier $|\Delta|$.
In general conditions, Hamiltonian~\eqref{Eq.FirstOrderHam} does not posses the chiral symmetry because of the effective on-site potential and neither the spatial inversion symmetry.
When all the tunneling processes are resonant, the effective on-site potential vanishes, and the Hamiltonian~\eqref{Eq.FirstOrderHam} becomes chiral symmetric.
In~\ref{AppB}, we derive the full conditions under which all the tunneling processes are resonant.

\section{Derivation of the full resonance condition}\label{AppB}
\begin{figure}[!bp]
\includegraphics[width=1\columnwidth]{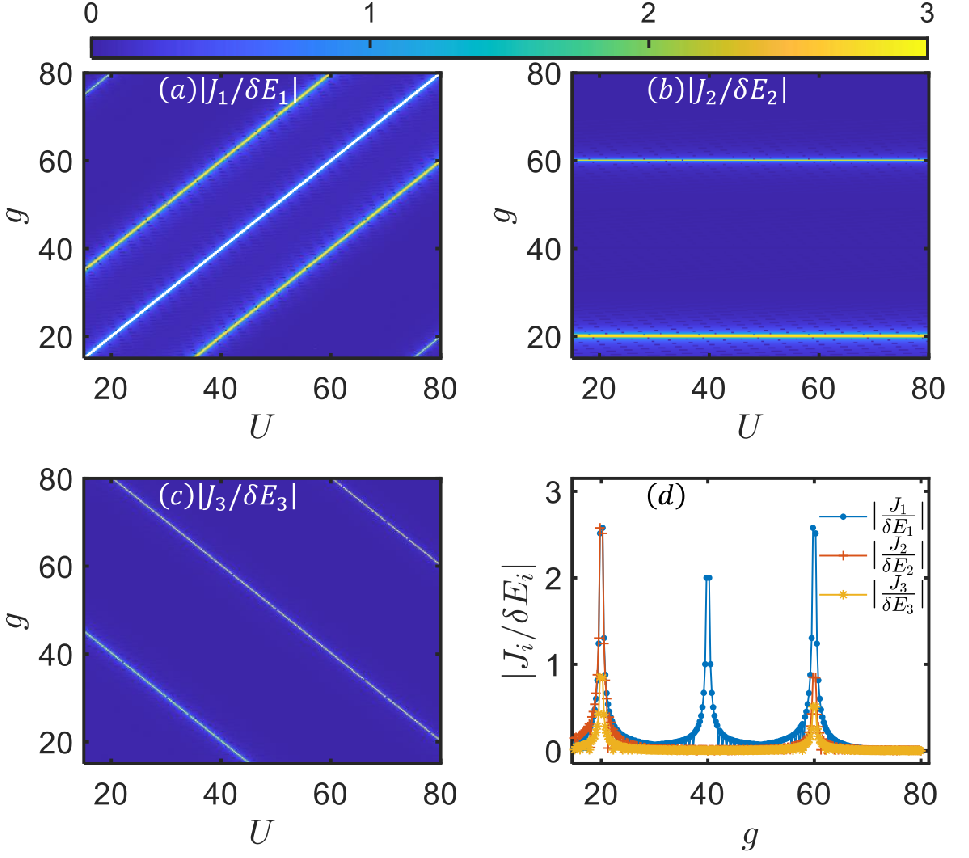}
  \caption{\label{Fig_J123}
  (a)-(c) $|J_i/\delta E_i|$ versus $g$ and $U$ at $\omega=20$ and $u=0.5$.
  The white regimes are the divergent points with $|J_i/\delta E_i|\to\infty$.
  (d) $|J_i/\delta E_i|$ versus $g$ at $U=40$, $\omega=20$ and $u=0.5$.
  The data at divergent points are not shown.
  }
\end{figure}

In general, a tunneling process is largely biased by the effective on-site potential.
We can derive the resonance condition for a tunneling process under which the effective on-site potential difference vanishes and the tunneling strength is nonzero.
Analytically, if $|\Delta|$ for a tunneling process is zero or can be compensated by the frequency provided by the driving, then the tunneling is resonant.
The driving strength $Ju(t)$ can be expanded in the Fourier series as $Ju(t)=2iuJ\sum_{q\in\mathrm{odd}}\frac{e^{iq\omega}}{q\pi}$ where $q$ are odd integers.
So, the resonance condition for a tunneling process with energy barrier $|\Delta|$ reads
\begin{equation}\label{Eq.rcond}
|\Delta|=0 \ \mathrm{or} \ (2k+1)\omega,
\end{equation}
where $k$ are non-negative integers.
Numerically, in Fig.~\ref{Fig_J123}(a)-(c), we show the ratios $|J_i/\delta E_i|$ ($i=1,2,3$) between the tunneling strength $J_i$ and the corresponding potential difference $\delta E_i$ of the three tunneling processes in the parameter plane $(U,g)$ at $\omega=20$ and $u=0.5$.
Here, $\delta E_i=\frac{i}{T}\ln e^{-i|\Delta_i|T}$ with $|\Delta_1|=|g-U|$, $|\Delta_2|=g$ and $|\Delta_3|=g+U$.
It is obvious that $|J_i/\delta E_i|$ almost equals 0 unless $g$ and $U$ satisfy the resonance condition~\eqref{Eq.rcond}.
In Fig.~\ref{Fig_J123}(d), we plot $|J_i/\delta E_i|$ as functions of $g$ along the cut $U=40$.
It is clear that $|J_i/\delta E_i|$ show peaks at $|\Delta_i|=0$ and $(2k+1)\omega$.

It is possible that the resonance conditions for the three tunneling processes are satisfied simultaneously.
To derive the full resonance condition, we set $U=ag$ with $a>0$, since we consider repulsive interaction.
The energy barriers then read $|\Delta_1|=|1-a|g$, $|\Delta_2|=g$ and $|\Delta_3|=(1+a)g$.
If $a=1$, it is obvious that the resonance conditions for the three processes can not be simultaneously satisfied.
For $a\ne1$, the full resonance condition is given by
\begin{eqnarray}\label{Eq.FRConditionFormula}
&&|1-a|g=(2k_1+1)\omega, \nonumber\\
&&g=(2k_2+1)\omega, \nonumber\\
&&(1+a)g=(2k_3+1)\omega,
\end{eqnarray}
with $k_{1,2,3}$ being non-negative integer numbers.
The solution of Eq.~\eqref{Eq.FRConditionFormula} reads
\begin{eqnarray}\label{Eq.FC1}
\frac{g}{\omega}&=&2k_2+1,\nonumber\\
k_3&=&2k_2-k_1, \nonumber\\
a&=&1-\frac{2k_1+1}{2k_2+1}, \nonumber\\
k_2&>&k_1,
\end{eqnarray}
if $a<1$, and
\begin{eqnarray}\label{Eq.FC2}
\frac{g}{\omega}&=&2k_2+1,\nonumber\\
k_3&=&2k_2+k_1+1, \nonumber\\
a&=&1+\frac{2k_1+1}{2k_2+1},
\end{eqnarray}
if $a>1$.
The combination of of Eq.~\eqref{Eq.FC1} and~\eqref{Eq.FC2} gives rise to the full resonance condition~\eqref{Eq.frescondition} in the main text.

\section{Derivation of the dimension difference between the chiral symmetric and anti-symmetric subspaces}\label{AppC}
Defining the dipole moment operator as $\hat D=\sum_jj\hat n_j$, the chiral operator reads $\hat{\mathcal C}=(-1)^{\hat D}$.
The dipole moments of a Fock state $|\mathbf{n}\rangle=|n_1n_2\cdots n_L\rangle$ and its spatial reversed state $|\bar{\mathbf{n}}\rangle=|n_L\cdots n_2n_1\rangle$ satisfy $D_{\mathbf n}+D_{\bar{\mathbf{n}}}=N(L+1)$, where $D_{\mathbf n}=\langle \mathbf n|\hat D|\mathbf{n}\rangle$.
If $N$ is odd, $D_{\mathbf{n}}$ and $D_{\bar{\mathbf{n}}}$ are of opposite odevity, which means that the Fock states in the chiral symmetric and anti-symmetric subspaces are in one-to-one correspondence through the spatial reversal transformation.
So, $N_+=N_-$.

When $N$ is even, $D_{\mathbf n}$ and $D_{\bar {\mathbf n}}$ are of the same odevity, which does not guarantee $N_+=N_-$.
Notice that one can write the dipole moment of a Fock state $|\mathbf{n}\rangle$ as $D_{\mathbf{n}}=\sum_{j\in \mathrm{odd}}jn_j+\sum_{j\in \mathrm{even}}jn_j$, where the first and last summations account for the summation over odd and even lattice sites, respectively.
Since the last term must be even, the odevity of $D_{\mathbf{n}}$ is decided by the first term.
If the total number of particles on the odd lattice sites is odd, $D_{\mathbf n}$ is odd; otherwise, it is even.
Thus, the total number of Fock states in the chiral symmetric subspace reads
\begin{equation}
N_+=\sum_{k=0}^{N/2}\binom{N}{2k}\binom{N}{N-2k}=\sum_{k=0}^{N/2}\left[\binom{N}{2k}\right]^2,
\end{equation}
where $\binom{x}{y}=\frac{x!}{y!(x-y)!}$ is the binomial coefficient and the expression in the summation is the number of ways to arrange $2k$ particles on the odd lattice sites and $N-2k$ particles on the even lattice sites.
Similarly, the total number of Fock states in the chiral anti-symmetric subspace reads
\begin{equation}
N_-=\sum_{k=0}^{\frac{N}{2}-1}\binom{N}{2k+1}\binom{N}{N-2k-1}=\sum_{k=0}^{\frac{N}{2}-1}\left[\binom{N}{2k+1}\right]^2.
\end{equation}
Using the equation $(1+x)^N(1-x)^N=(1-x^2)^N$ and performing the binominal expansion of both sides, one can obtain
\begin{equation}
\sum_{k=0}^{N}(-1)^k\left[\binom{N}{k}\right]^2=(-1)^{N/2}\binom{N}{N/2},
\end{equation}
which means $N_+-N_-=(-1)^{N/2}\binom{N}{N/2}$.

We notice that the dipole moment of the pinnacle state reads $\frac{N(N+1)}{2}$.
When $N$ is even, its odevity is decided by $N/2$, so the pinnacle state is always on the subspace with larger dimension.

\section{Finite-size effect}\label{AppD}
\begin{figure}[!tbp]
\includegraphics[width=1\columnwidth]{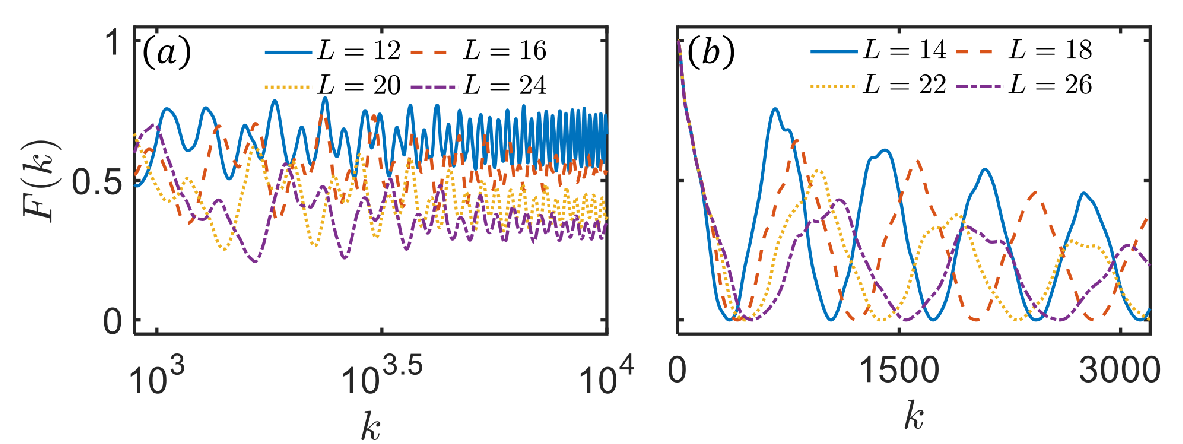}
  \caption{\label{Fig_dynamics_moreL}
  The fidelity $F(k)$ versus $k$ after quenching from $|\mathrm{tp}\rangle$ at system sizes (a) $L=12$, 16, 20 and 24 and (b) $L=14$, 18, 22 and 26.
  The initial state is $|\mathrm{tp}\rangle$.
  The other parameters are $U/g=2$, $g/\omega=1$, $g=50$ and $u=0.5$.
  }
\end{figure}

\begin{figure}[t]
\includegraphics[width=1\columnwidth]{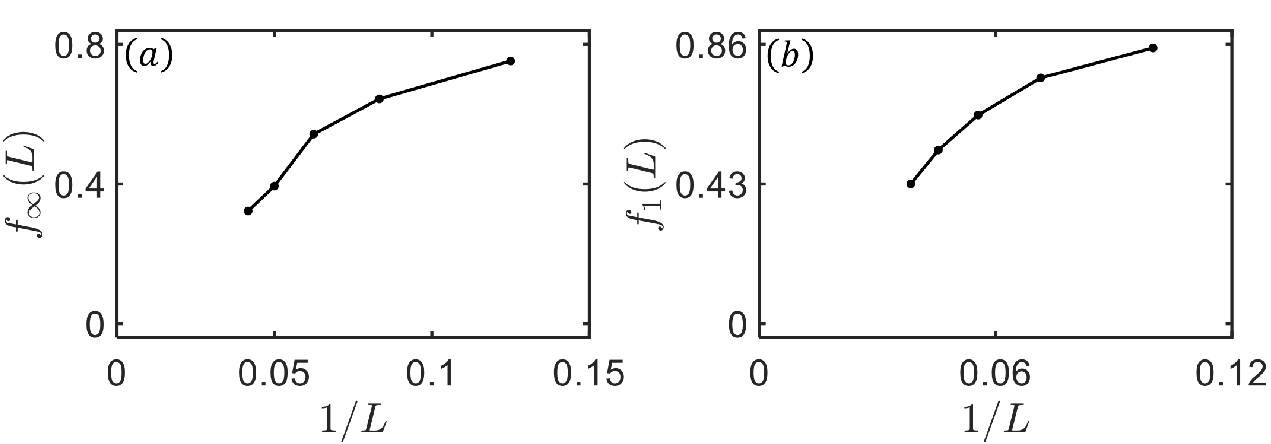}
  \caption{\label{Fig_scalingF}
  (a) The infinite time average of the fidelity $f_\infty$ versus $1/L$ for $L=$8,12,16,20 and 24.
  (b) The fidelity at the first revival peak $f_1$ versus $1/L$ for $L=$10,14,18,22 and 26.
  The initial state is $|\mathrm{tp}\rangle$.
  The other parameters are $U/g=2$, $g/\omega=1$, $g=50$ and $u=0.5$.
  }
\end{figure}

We now analyze the finite-size effect on the scarred dynamics.
For this purpose, we simulate the evolution for more system sizes.
In Fig.~\ref{Fig_dynamics_moreL}, we show the evolution of the fidelity from $|\mathrm{tp}\rangle$ for $L=12$, 14, $\cdots$, 26.
The evolution is performed through the exact diagonalization method for $L\leq 18$ and the Krylov space-based algorithm~\cite{Saad1992} for $L>18$.

For even particle numbers, the fidelity oscillates around fixed values $f_{\infty}(L)$, which decays as $L$ increases, see Fig.~\ref{Fig_dynamics_moreL}(a).
For odd particle numbers, the fidelity exhibits decaying collapse and revival, with the revival peaks decaying as $L$ increases, see Fig.~\ref{Fig_dynamics_moreL}(b).
We perform finite-size scaling of $f_{\infty}(L)$ for even particle numbers and the first revival peak $f_1(L)$ for odd particle numbers.
$f_{\infty}(L)$ is obtained by averaging the fidelity over driving cycles $k\in[9000, 10000]$.
Both $f_{\infty}(L)$ and $f_1(L)$ decrease and show tendency to 0 as $L$ increases, see Fig.~\ref{Fig_scalingF}.
However, we can not obtain analytical formulas of the scaling law of them, because the number of accessible system sizes is limited.
We expect that $f_{\infty}(L)$ and $f_1(L)$ will tend to zero in the thermodynamic limit from the following consideration.
The scarred dynamics comes from the interference effect of single particle walking in the tower and the suppression of wave function from leaking from the tower.
The size of the tower is $L$. As $L\to \infty$, the particle can never reach the boundary of the tower and the interference effect can never show.
In addition, the number of edges connecting the tower and the thermal part is $L-1$, which will further enhance the leakage of the wave function from the tower as $L$ increases.
So, we expect that $f_{\infty}(L)$ and $f_1(L)$ decay to 0 and thus the scarred dynamics does not persist in the thermodynamic limit.
Rigorous numerical demonstration deserves further study.
We note that the behavior $f_1$ decaying to 0 is also observed in other models with typical scarred dynamics~\cite{Desaules2021,Guo2023}.

\begin{figure}[!htbp]
\includegraphics[width=1\columnwidth]{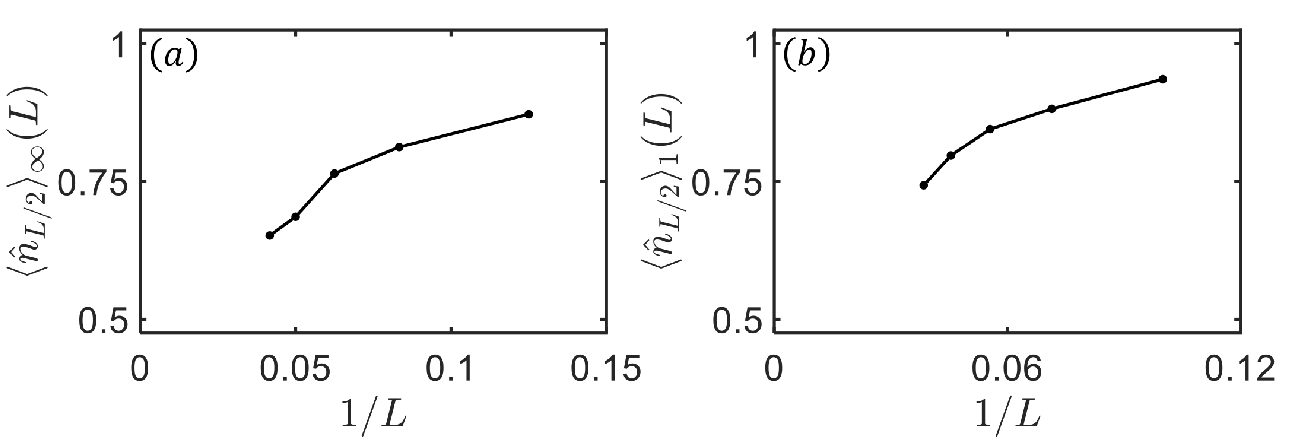}
  \caption{\label{Fig_scaling_NN}
  (a) $\langle \hat n_{L/2}\rangle_{\infty}(L)$ versus $1/L$ for $L=$8,12,16,20 and 24.
  $\langle \hat n_{L/2}\rangle_{\infty}(L)$ is obtained by averaging $\langle \hat n_{L/2}\rangle$ over driving cycles $k\in[9000,10000]$.
  (b) $\langle \hat n_{L/2}\rangle_1$ versus $1/L$ for $L=$10,14,18,22 and 26.
  The initial state is $|\mathrm{tp}\rangle$.
  The other parameters are $U/g=2$, $g/\omega=1$, $g=50$ and $u=0.5$.
  }
\end{figure}

The finite-size effect on the scarred dynamics is also represented by the dynamics of the local observables.
For even $N$, we find that the local density $\langle \hat n_{L/2}\rangle$ oscillates around fixed values $\langle \hat n_{L/2}\rangle_{\infty}(L)$, which decreases as $L$ increases (results not shown here).
For odd $N$, the local density $\langle \hat n_{L/2}\rangle$ shows collapse and revival, with the revival peak decaying as $L$ increases (results not shown here).
In Fig.~\ref{Fig_scaling_NN}, we show $\langle \hat n_{L/2}\rangle_{\infty}$ (for even $N$) and the first revival peak $\langle \hat n_{L/2}\rangle_1$ (for odd $N$) as functions of $1/L$.
Both of them decay as $L$ increases and show tendency to the thermal value $0.5$ in the thermodynamic limit.
It means that the even-odd effect might not be observed in the thermodynamic limit.


%

\end{document}